\documentclass{article}

\usepackage[utf8]{inputenc}
\usepackage{titlesec}
\usepackage{multirow}
\usepackage{pdfpages}
\usepackage{authblk}

\title{Malware Dynamic Analysis Evasion Techniques: A Survey}

\author[1]{Amir Afianian}
\author[1]{Salman Niksefat}
\author[1]{Babak Sadeghiyan}
\author[2]{David Baptiste}
\affil[1]{APA Research Center, Amirkabir University of Technology}
\affil[2]{ESIEA (C + V)O lab}
\affil[]{\textit{\{a.afianian, niksefat, basadegh\}@aut.ac.ir}}
\affil[2]{\textit{baptiste.david@esiea.fr}}

\date{June 2018}

\begin{document}

\maketitle

\begin{abstract}
The Cyber world is plagued with ever-evolving malware that readily infiltrates all defense mechanisms, operates viciously unbeknownst to the user and surreptitiously exfiltrate sensitive data. Understanding the inner workings of such malware provides a leverage to effectively combat them. This understanding, is pursued through dynamic analysis which is conducted manually or automatically. Malware authors accordingly, have devised and advanced evasion techniques to thwart or evade these analyses. In this paper, we present a comprehensive survey on malware dynamic analysis evasion techniques. In addition, we propose a detailed classification of these techniques and further demonstrate how their efficacy hold against different types of detection and analysis approach. 

Our observations attest that evasive behavior is mostly interested in detecting and evading sandboxes. The primary tactic of such malware we argue, is fingerprinting followed by new trends for reverse Turing test tactic which aims at detecting human interaction.
Furthermore, we will posit that the current defensive strategies beginning with reactive methods to endeavors for more transparent analysis systems, are readily foiled by zero-day fingerprinting techniques or other evasion tactics such as stalling. Accordingly, we would recommend pursuit of more generic defensive strategies with emphasis on path exploration techniques that has the potential to thwart all the evasive tactics. 
\end{abstract}

\section{Introduction}
Gaining access through ARPANET network, Creeper copied itself remotely to other computers, and, prompted the message: “Catch me if you can”. Emerged as an experiment in 1971 \cite{Russell1991}, \textit{Creeper} carried no harmful purpose, yet, with its quick propagation it cast the light on a future with individuals or entities incorporating their malicious intentions into a software (hence the name malware), that are capable of inducing undesired and harmful effects on the infected systems \cite{Bayer2006a}. In those days, developing malware was merely about show off; but, in today’s world, as the Internet has integrated every individual, institute, and organization into a single cohesive complex, the goal of malware authors have extended to include far more lucrative objectives, i.e. money, intelligence, and power. Accordingly, these possibilities have profoundly motivated a new malware industry, the outcome of which, are highly sophisticated variants, which aptly infiltrate systems and operate viciously unbeknownst to the user or defense mechanisms.

Devising countermeasure or technologies that can withstand this level of sophistication would be possible merely by understanding the precise inner workings of such malware. Malware analysis is the way to achieve this understanding \cite{Distler2007}. Initially, with the help of disassemblers, decompilers etc. analysts inspected the malware’s binary and code to infer its functionality. This approach, which is also referred to as static analysis became far more arduous, and intricate with the advent, and evolution of code obfuscation tactics \cite{Rad2012,Schiffman2010,Schiffman2010a,You,Borello2008} and other evasion tactics targeting static analysis e.g. opaque constants \cite{Moser}, packers \cite{Guo}, etc. As a resolution, a promising approach that was adopted was dynamic analysis in which the basis of analysis and detection is what the file does (behavior) rather than what the file is (binary and signature) \cite{Mourad2015}. Put differently, in dynamic analysis, an instance of the suspected program is run and its behavior is inspected in run-time. This approach would obviate the hurdles posed by aforementioned static analysis evasion tactics. To thwart these efforts, however, malware authors turned to a new category of evasion tactics that targeted dynamic analysis. 

In this paper, we identify two modes of dynamic analysis i.e. manual and automated. Manual dynamic analysis is a more traditional form of dynamic analysis and is often conducted with the help of debuggers. Automated dynamic analysis is a more novel approach and also a response to the ever-increasing new samples that security vendors face on a daily basis. Automated dynamic analysis is often represented by the Sandbox technology. Further, we conduct a comprehensive survey on evasion techniques that are tailored to each mode of analysis. 

In addition, we will portray the current and also the emerging evasive trends. In the case of manual dynamic analysis evasion we posit that anti-debugging techniques are still significantly being practiced. Also, we will discuss how the new fileless malware is emerging and prevailing the traditional dynamic analysis. Moreover, in the case of automated dynamic analysis, we divide evasion tactics in two broad categories: detection-dependent, and detection-independent. We argue that until recent times, sandbox technology have been focusing on technologies that chiefly target the detection-depend evasion tactics. As a response, malware authors are gradually adopting more and more detection-independent tactics to evade such automated dynamic analysis environments.

Our goal in this paper is threefold. First, given the importance of the subject, we have aimed at offering a comprehensive classification, with the hope of moving towards an established taxonomy. Second, we aspired to unveil the direction toward which the evasive behavior is trending. Thus, we tried to yield a view of the current situation for pervasiveness of each evasion tactic. Our third goal is to identify the shortcomings of the current defensive strategies and offer a direction which we deem to have more potential on effectively confront evasive malware.

Our contributions are:
\begin{itemize}
  \item We present a comprehensive survey of malware dynamic analysis evasion techniques for both modes of manual and automated. There have been several surveys on malware analysis evasion such as \cite{bulazel2017survey} which has merely focused on automated dynamic analysis evasion and covers only one (out of five that we have identified) evasion tactic. Other surveys like \cite{gao2014survey, marpaung2012survey} trivially surveyed analysis evasion and provide no detailed overview of malware dynamic analysis evasion. 
  \item For both manual and automated modes, we present a detailed classification of malware evasion tactics and techniques. To the best of our knowledge, this would be the first comprehensive survey of dynamic analysis evasion tactics that offers a thorough classification. 
  \item We portray the current trends in the realm automated dynamic analysis evasion techniques and discuss about the proper course of action for the future directions. 
\end{itemize}

\subsection{Overview} 
\subsubsection{Scope}Latest reports attest to the dominance of Windows malware by a staggering ratio of 77.22\% in 2017 \cite{av-test2017}. Statistics also show that due to the employment of evasion techniques,  “64\% of AV scanners fail to identify 1\% of the malware after 1 year" \cite{Kruegela}. These figures have further motivated us to converge our focus on evasive malware targeting Windows OS.  Moreover, in our research, we have strove to probe the intersection of academia and industry. Thus, we have reviewed both academic papers, and related industry-provided literature.
\subsubsection{Approach} Our primary endeavor has been to provide a detailed and vivid overview of dynamic analysis evasion techniques. To this end, we have classified the results of our scrutiny in three levels: category, tactic, and technique. Each category represents a particular approach and a goal which is both shared and pursued through different tactics. Each tactic then, is implemented via different techniques. Throughout the survey we elaborate the details down to the tactic level and we discuss several representative techniques. At the end of each section we provide a summarizing table.
\subsubsection{Evaluation} To fulfill the goals of this survey, we have employed several criteria such as efficacy and pervasiveness of different techniques which serve as the basis of comparisons for the related topic. In our summarizing tables we present data from reviewed works where statistics are available, and otherwise we offer evaluations based on our own observation.
\subsubsection{Organization} In section II, We discuss manual dynamic analysis (debugging) and corresponding anti-debugging tactics. In section III we discuss automated dynamic analysis and the corresponding evasion tactics. section II and III are followed by a discussion. Finally, section IV concludes the paper. 
Figure 1 demonstrates our proposed classification of the malware dynamic analysis evasion tactics.
\begin{figure}[h!]
  \caption{A classification of malware dynamic analysis evasion tactics}
  \includegraphics[scale=0.25]{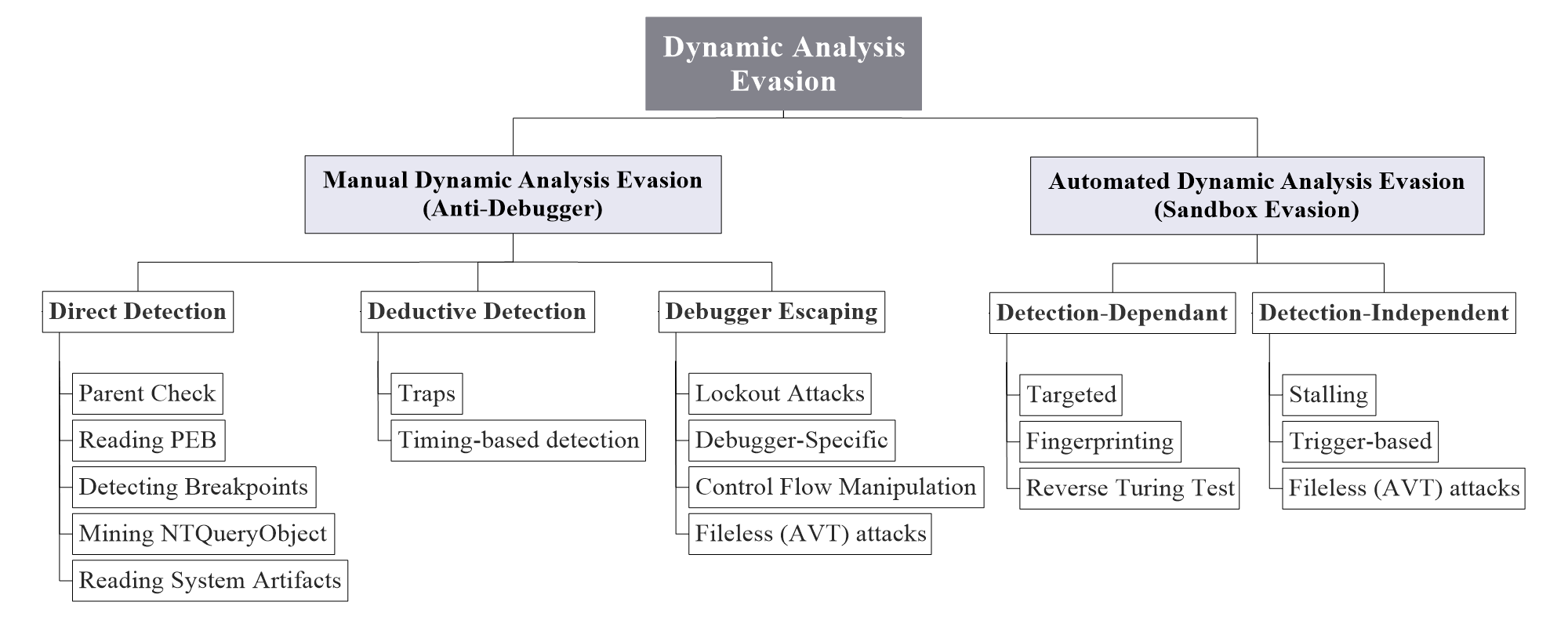}
\end{figure}

\section{Preliminaries}
\subsection{Evasion and Transparency}
In literature, evasion constitutes a series of techniques employed by malware in order to remain stealth, avoid detection, or hinder efforts for analysis. A major evasion tactic as we will discuss, is fingerprinting \cite{oyama2018trends}. With fingerprinting, malware tries to detect its environment and verify if it is residing in a production system or an analysis system.

In the same level, one major strategy is to hide the cues and clues that might expose the analysis system. A system is more transparent if it exposes fewer clues to malware \cite{Kruegel2014}.
\subsection{Manual vs Automated}
Manual and automated analysis are two major terms which form the basis of our classifications. The manual analysis is when the analysis process is performed by an expert human. Automated, on the other hand, is the process that is performed automatically by a machine or software. 
\subsection{Detection vs Analysis}
Previously, there was no need for defining these two terms. Detection would simply refer to the process of discerning if a given file is malicious or not; analysis, on the other hand, would refer to the process of understanding how the given malware work. Today, however, this dividing line is blurry. The reason is that the role of automated analysis tools such as sandboxes is now extended. In addition to reporting on malware behavior, sandboxes are now playing their role as the core of automated detection mechanisms \cite{Kruegel2014}. In this paper, we follow the similar concepts when uttering "analysis". For "manual" it would mean understanding the malware behavior \cite{Sikorski2012}, for the automated, additionally it can mean detection.
\subsection{Static vs Dynamic}
There are two major types of analysis i.e. static, and dynamic. Static analysis is the process of analyzing the code or binary without executing it. Dynamic analysis is the process of studying the behavior of the malware (API, system calls, etc.) at run-time. Both types of the analysis can be performed either manually or automatically. 
In this paper, our focus is on dynamic analysis and how malware tries to prevent or evade such analyses.
\subsection{Category, Tactic, and Technique}
Throughout this paper, we use the terms "category", "tactic", and "technique" that are the basis of our classification. Category of evasion is our high-level classification. Each category has the goal of evasion with a specific attitude for achieving it. This attitude is highly correlated with the efficacy of the evasion and is pursued by the tactics under each category. Tactics, in other words, are the specific maneuvers or approach for evasion with the specified attitude of its parent category.
Finally, techniques are the various practical implementation of those tactics. 

\section{Manual Dynamic Analysis Evasion}
As cited in the introduction, due to the employment of code obfuscation, packers, etc. static analysis of malware has become a daunting task. To obviate the conundrums and limitations of this approach, analysts opt for dynamic analysis in which malware’s behavior is inspected in run-time and often with the help of debuggers. This way of examination has two major benefits; one is that it relieves us from the impediments inflicted by packers, polymorphism, etc. Second, taking this approach enables us to explore the activities that manifest themselves only in runtime \cite{Quynh2010} such as the interaction of the program with the OS \cite{Mourad2015}. We view this approach which is aided by debuggers, under the term “manual dynamic analysis”. The corresponding evasion tactics to this approach involves the set of employed techniques within malware code with the goal of hindering, impeding or evading the analysis process. These measures include approaches such as detecting the presence of analysis tools on the system (e.g. wireshark, tcpdump, etc.) or detecting virtual-machines as a sign of analysis environment,  but, the majority of manual analysis evasion techniques are targeted toward debuggers which are the primary tools of manual dynamic analysis. Hence, in the domain of manual dynamic analysis evasion, we converge our focus on anti-debugging tactics.

Simply, anti-debugging is the application of one or more techniques, to fend off, impede or evade the process of manual dynamic analysis (debugging) or reverse-engineering. It is noteworthy to mention that anti-debugging techniques, similar to obfuscation, are legitimate practices conducted by developers to protect their software. This implies that anti-debugging has a long history which is followed by a wide and diverse set of techniques. In this section, we will present our survey on the most cited manual dynamic analysis evasion (anti-debugging) tactics and corresponding techniques \cite{Mourad2015, Quynh2010, Branco2012,apriorit2016a, Chen, falliere2007, Ferrie2011,Sikorski2012,Tully9Nov2008,Yason2007,Zhang} that are more relevant to the context of malware. In addition, for each technique, we employ 4 criteria to better yield an overview of relative weight and efficacy. We summarize our anti-debugging study in table 1. Three criteria for our comparison and our touchstone for evaluating them are as follows:

\begin{itemize}
    \item \textbf{Complexity:} The difficulty of incorporating and implementing the technique within malware code. We associate low complexity if the technique involves simple API calls for detection of debuggers’ presence, medium to those that require several calls to inquire information from the system as well and the need to deduce from the returned information the probability of debuggers presence, and high to those that are challenging to implement and require more than 30 lines of assembly code.
    
    \item \textbf{Resistance:} Resistance pertains to the difficulty level of counteracting the evasion technique. If it would be as simple as altering some system information (e.g. randomizing registry key or a bit in \textit{PEB}) we consider it low; if defeating the anti-debugging technique requires user-level hook and \textit{dll} injection we assign medium resistance; if kernel-level manipulations e.g. drivers is needed we set the resistance level as high; and if the countermeasure requires operations below OS (virtualization, emulation, etc.) we consider it very-high. Finally, we have N/A, associated to techniques for which there exist no countermeasures yet. 

    \item \textbf{Pervasiveness:} Due to the extreme diversity of malware sample dataset it is hard to establish a ground truth for the pervasiveness of each tactic. However, we strive to provide a fair view of this metric based on other works \cite{Branco2012,Shi} in addition to our own observations and experience in the field.  
\end{itemize}
Additionally, we allot a column to textbf{Countermeasure} under Which we briefly note how the anti-debugging technique could be circumvented.

 Before scrutinizing the anti-debugging techniques, we will have a brief introduction on debuggers, their goals, types, and techniques. Building upon this understanding, we can better understand and elaborate anti-debugging tactics and what makes them feasible.
 
 \subsection{A Briefing on Debuggers}
 As Sikorski states, “A debugger is a piece of software or hardware used to test or examine the execution of another program \cite{Sikorski2012}”.  Among the core functionality of debuggers \cite{Aggarwal2002}, there are several features that are critical to malware analysis such as stepping through the code one instruction at a time, pausing or halting it on desired points, examining the variables, etc. To provide each of the mentioned functionality, debuggers rely on different tactics and each tactic is often aided with specific hardware or software provisions that inevitably result in subtle changes on the system. For instance, to provide the pausing capability, one of debuggers’ tactics is to set breakpoints. Using breakpoints is further assisted with either special hardware \cite{Chourdakis2008} (e.g. DR Registers of CPU and specific API to access/alter them), or software (e.g. specific opcodes, etc.) \cite{msdn} mechanisms. Single stepping, as another instance, is made possible by triggering exceptions in the code which is aided by a specific flag (trap flag). One way a malware can discover the presence of the debugger is through the very same aiding mechanisms or results of corresponding changes on the system which we will elaborate more as we proceed. 
 
 Based on the needed functionality, debuggers are implemented following different approaches \cite{Gao}, including user-mode debuggers e.g. \textit{OllyICE}, \textit{OllyDbg}, kernel debuggers e.g. \textit{WinDbg}, KD, and simulator-based debuggers e.g. \textit{Ether} \cite{Dinaburga}, \textit{BOCHS} \cite{Lawton2003}. Each of the designs, yield different levels of transparency. Kernel-level debuggers are more transparent than User-level debugger and provide more detailed information as they operate in ring 0 (same level of privilege as the OS). Simulator-based debuggers have an even higher level of privilege than kernel debuggers since the OS in this setting is running atop the simulated (virtualized/emulated) hardware \cite{Dinaburga, Lawton2003} and consequently are more transparent to malware. In the next section, we proceed to manual dynamic analysis evasion, or more specifically, anti-debugging tactics.

 \subsection{Proposed Anti-Debugging Classification}
 We propose three major categories, as malware’s initial anti-debugging strategies: direct detection, deductive detection, and debugger evasion. As stated earlier, we elaborate on each category down to the tactic level along with a couple of representative techniques for each tactic. Finally, we summarize the results of our scrutiny in table 1.
 
 \subsubsection{\textbf{Direct Detection}}  We call the tactics of this category as “direct detection” in that, the malware, simply through direct observation of debugger’s byproducts, and artifacts. can spot its presence. One of the malware’s major leverages against debuggers stems from the fact that debuggers were originally devised to debug legitimate software \cite{Quynh2010}. Thus, initially, there has been no stealthy countermeasure provisioned by them. Moreover, when we want to enable debugging capability, we have to instrument the system with necessary tools which obviously results in a wide spectrum of traces in different levels of the system \cite{Chen, Shi}. Examining the system for such traces that are inherent to working principles of debuggers, is one major strategy that malware utilizes to detect the presence of a debugger. In this section, we will elaborate on these system changes and how malware might try to disclose them. This category of tactics are often easy to employ and also easy to defeat. Moving from user-mode debuggers to simulation-based ones, this category of evasion pales in efficacy.
\paragraph{\textbf{Reading PEB}} Process Environment Block (\textit{PEB}) is a data structure that exists per process in the system and contains data about that process \cite{Microsoft}. Different sections of \textit{PEB} contain information that can be probed by a malware to detect whether a debugger exists. The most obvious one is a field inside PEB named \textit{BeingDebugged} which can be read directly or as Microsoft recommends— and malware like \textit{Kronos} \cite{Hub2014} or \textit{Satan RaaS} \cite{Team2017} implement — through the specific APIs that read this field, i.e. \textit{IsDebuggerPresent()}, \textit{CheckRemoteDebuggerPresent()}. Implementation of this technique is of trivial complexity which can be countered through alteration of BeingDebugged bit \cite{Oleg2016} or API hook \cite{apriorit2016a}. Other techniques based upon \textit{PEB} involve utilization of \textit{NtGlobalFLags} and \textit{ProcessHeap} which are slightly more complex \cite{Ferrie2011}. Collectively, anti-debugging tactics relying on \textit{PEB}, constitute the majority of anti-debugging techniques observed in malware \cite{Branco2012}.  

\paragraph{\textbf{Search for Breakpoints}} To halt the execution, debuggers set breakpoints. This can be accomplished through hardware or software techniques. In hardware breakpoints, the breakpoint address, for instance, can be saved in CPU DR registers or. In software breakpoints, the debugger would write the special opcode \textit{0xCC} (\textit{INT 3} instruction) into the process which is specifically designated for setting breakpoints. Consequently, the malware, if spots signs of these breakpoints, presumes the presence of a debugger. This can be accomplished through a self-scan or integrity check, looking for \textit{0xCC}, or using \textit{GetThreadContext} to check CPU register \cite{Ferrie2011}. The latter technique has been employed by malware such as \textit{CIH} \cite{McAfee2000} or \textit{MyDoom}.

This tactic is the second most observed anti-debugging technique in malware’s arsenal \cite{Branco2012} and is simple to implement (less than 10 lines of assembly code often suffices). Countering this category of tactics, however, is not trivial. In the case of software breakpoints, for instance, the debugger has to keep and feed a copy of the original byte that was replaced by \textit{0xCC }opcode \cite{Oleg2016}.

\paragraph{\textbf{System Artifacts}} From installation to configuration and execution, debuggers leave  traces behind in different levels of the OS, e.g. in the file system, registry, process name, etc. Hence, a malware can simply look for these traces.\textit{ FindWindow()}, \textit{FindProcess()} are a couple of APIs that \textit{shcndhss} \cite{JOESandbox} exploited to detect debuggers. The malware, for instance can give the name of debuggers as the parameter to \textit{FindWindow()} to verify if its process exists on the system or not \cite{Tully9Nov2008, shields2010anti}. Most often, simple anti-debugging techniques are trivially defeated. The countermeasure to this category of tactic is randomizing the names, or altering the results of the aforementioned query through simple API hooks. Even though attributed to one of the anti-debugging techniques in literature, we have rarely observed it in the wild.

\paragraph{\textbf{Parent Check}} Ordinarily, applications are executed either through double clicking of an icon, or execution from command line, the parent process ID of which is retrievable accordingly. It would be a pronounced sign of debugger if the examined parent process name belongs to a debugger or is not equivalent to process name of \textit{explorer.exe} \cite{Ferrie2011}. One straight forward technique is using CreateToolhelp32Snapshot() and checking if the parent process name matches the name of a known debugger \cite{apriorit2016a}. A randomization of debuggers' process name suffices to defeat this technique. Another more complex way is to check the parent process ID against that of explorer.exe. A malware such as \cite{JOESandbox} utilizes of the following functions: \textit{GetCurrentProcessId()}, \textit{CreateToolhelp32Snapshot()}, \textit{Process32First()} \textit{Process32Next()} \cite{Branco2012}. If the parent process ID does not match with that of the \textit{explorer.exe}, it might be a sign of debuggers presence. Precise and effective employment of this technique faces some challenges. If there are more than one process for \textit{explorer.exe}, then, there would be more than one PID for it. This would complicate the process by sparking false positives in that the parent PID might match to one \textit{explorer.exe} PID and not the other. Countering this tactic would be skipping the relevant APIs \cite{Shi}. Our observations demonstrates few utilization of this tactic \cite{Branco2012}. 

\paragraph{\textbf{Mining NtQuerySystemInformation}} The \textit{ntdll} \textit{NtQuerySystemInformation()} is a function that accepts a parameter which is the class of information to query \cite{Ferrie2011}. Most of these classes are not documented. There are many ways malware might pursue to leverage this function with the goal of spotting debugger’s signs. Debugger is easily exposed, for instance, if the malware passes the \textit{SystemKernelDebuggerInformation()} (0x23) class as the parameter to this function, the returned values are two flags; one of them is \textit{KdDebuggerNotPresent()}, if the value equals zero, it attests debuggers presence. Employment of this tactic is not simple (more than 40 lines of assembly code \cite{Ferrie2011}), and nor is circumventing it. The reason is that when this function is invoked, the returned values are coming from the kernel. Defeating anti-debugging techniques following this tactic requires patching the kernel. \textit{vti-rescan} \cite{Virustotal2015}, \textit{Wdf01000.sys} \cite{JOESandbox2010}, and \textit{Inkasso trojaner} \cite{Cursec2013} are some instances that leverage this tactic. 

As stated earlier, deployment of direct detection tactics is relatively straightforward and trivial which in turn are often just as trivial to defeat.  more subtle way of detecting and evading debuggers is through deductive detection which is the subject of the next section

\subsubsection{Deductive Detection}
Tactics under this category fulfill the goal of inferring debuggers presence, rather than directly observing it. Tactics of this category calculate and assess the probability of debugger’s presence. These inferences mostly rely on exploiting the logic of the system (e.g. \textit{SEH} Exception handling \cite{Microsoft2018a}). More specifically, as a way of illustration, the way Windows handles exceptions can be abused by malware to illicit debugger presence). Malware may employ different ingenious tactics to this end which we explore in the following. 

\paragraph{\textbf{Traps}} Not to be mistaken with the trap flags, we define this category of tactics as “traps”. Following this tactic, the malware provisions codes that when traversed or stepped through by debugger, production of specific information or lack thereof, would help the malware deduce the presence of a debugger. Many techniques fit this category and we elaborate a couple of them here and hint at the rest in table 1. 

One technique by which malware beguile debuggers to disclose cues of its presence is using specific instructions and exploiting the logic of how these instructions are handled. The handling is performed through Structured Exception Handling \textit{SEH} (You can refer to\cite{Microsoft2018a} for more information about \textit{SEH}). For instance,  Max++ malware \cite{Institute2015}  embeds “\textit{int 2dh}” instruction within its code. According to \textit{SEH}, when this instruction is executed, in normal situation i.e. absence of debugger, an exception is raised and malware can handle it via a try-catch structure. However, if a debugger is attached, this exception will be transferred to the debugger rather than the malware; the absence of expected exception is the logic that the malware entertains to deduce the presence of a debugger. Malware may employ other techniques to lay their traps such as embedding specific instruction prefixes \cite{Tully9Nov2008}, or other  instructions such as 41h \cite{Ferrie2011}. These techniques are of low complexity and can be implemented with less than 20 lines of code (in assembly). One way of countering these traps requires debuggers to skip these instructions. According to our survey, utilizing traps is a fairly common approach \cite{Branco2012}. 

\paragraph{\textbf{Timing-Based Detection}} Timing-based detection are among the most efficacious tactics of inferring debugger’s presence. Adroit employment of this tactic reliably exposes the presence of a debugger and circumventing them is an arduous task. 

The logic behind this tactic follows malware authors’ presumption that a particular function or instruction set, require merely a minuscule amount of time. Thus, if a predefined threshold is surpassed, malware would infer the presence of a debugger. Timing-based detection can be carried out either locally with the aid of local APIs(\textit{GetTickCount()}, \textit{QueryPerformanceCounter()}, etc. \cite{RN79}) or CPU \textit{rdtsc} (read time stamp counter) or can be performed by inquiring an external resource through the network \cite{Pek} to conduct the timing. Local timing is simple to employ and difficult to circumvent. Countering timing-based detection conducted with the aid of external resource (using NTP or tunneled NTP), however, is still an open problem \cite{Zhang}.\textit{ W32/HIV} \cite{Kaspersky2000}, \textit{W32/MyDoom} \cite{McAfee2007}, \textit{W32/Ratos} \cite{McGraw2000} are infamous malware that are known to have exploited the timing discrepancies.

\subsubsection{Debugger Escaping} Unlike previous categories the goal of which, were detecting or inferring debugger’s presence, the goal of this category is assuring that the execution of malware wholly or partially takes place. Malware adopt catching approaches to fulfill the goal of this strategy. We will have a survey on tactics of this category in the following. 
\paragraph{\textbf{Control Flow Manipulation }} Through this tactic, malware exploit the implicit flow control mechanism conducted by Windows \textit{OS}. To implement these techniques, malware authors often rely on callbacks, enumeration functions, thread local storage (\textit{TLS}), etc. \cite{Ferrie2011, Shi}. There are several noteworthy techniques that fulfill the goals of “debugger Escaping” and each deserves a brief introduction. 

\begin{itemize}
    \item \textit{Thread-hiding}:A simple and effective technique is thread-hiding which if used, prevents debugging events from reaching debugger. Microsoft has provisioned special API \cite{msdn2018, WalterTiezhuKong2013}. This technique uses \textit{NtSetInformationThread()} function to set the field \textit{HideThreadFromDebugger()} of ETHREAD kernel structure \cite{Yason2007}. This is a powerful technique and simple to implement and can be countered by hooking the involving functions. \textit{LockScreen} \cite{WalterTiezhuKong2013} is an instance known to have utilized this technique.
    \item \textit{Suspending Threads}:  A more aggressive way malware might step into is striving to halt the process of the debugger to continue its own execution with little trouble. This technique can be effective against only user-mode debuggers and can be carried out by leveraging \textit{SuspendThread()} or \textit{NtSuspendThread()} from \textit{ntdll} \cite{Branco2012}. Suspending threads is one of the anti-debugging techniques that \textit{Kronos} banking malware used in its arsenal \cite{Hub2014}. 
    \item \textit{Multi-threading}: Another technique to bypass debuggers and continue the execution is multi-threading. One way to implement it through \textit{CreateThread()} API. Malware that is packed, often spawns a separate thread within their process to perform the decryption routines \cite{Yason2007}. However, there are instances where malware executes a part of its malicious code through a different thread outside the debugger. \textit{McRat} \cite{UIC2013} and \textit{Vertexnet} \cite{CTurt} have incorporated such technique. Countering this technique is tricky. One way is to set breakpoints at every entry point \cite{Monoxide2016, Ferrie2008}.  
    \item \textit{Self-debugging}: Self-debugging is an interesting technique which prevents the debugger from successfully attaching to the malware \cite{XPN2017}. By default, each process can be attached to merely one debugger. Malware exploits this by running a copy of itself and attaches to it as a debugger. Hence preventing another debugger to own it. There are several ways to implement this tactic, for instance by leveraging \textit{DbgUiDebugActiveProcess()} or \textit{NtDebugActiveProcess()}.
\end{itemize}
Collectively, debugger escaping techniques are not much common among the samples we have observed. 
\paragraph{\textbf{Lockout evasion}} In Lockout tactic, malware accomplishes the goal of continuing its execution by impeding with the working of the debugger. Malware achieve this through several techniques. One is to opt for \textit{BlockInput()} function as in the case of \textit{Satan RaaS} \cite{Team2017}, through which malware prevents mouse and keyboard inputs until its conditions are satisfied. Other techniques involve exploiting a feature in Windows NT-based platforms that allow for the existence of multiple desktops. Malware, such as \textit{LockScreen}, \cite{WalterTiezhuKong2013} with the help of \textit{CreateDesktop()} followed by \textit{SwitchDesktop()} can select a different active desktop and continue its working unbeknownst to the debugger \cite{Ferrie2008}.

\paragraph{\textbf{Debugger Specific}} Debugger specific evasion exploit vulnerabilities that are exclusive to a specific debugger. These vulnerabilities are difficult to discover, but simple to put into action. A famous instance pertains to \textit{OllyDBG} \cite{Yason2007}. Regular versions of this debugger have a format string bug which can be exploited to cause it to crash by passing an improper parameter to \textit{OutPutDebugString()} function. Another pervasive-at-the-time technique was related to \textit{SoftICE} debugger which was susceptible to multiple DoS attacks due to two vulnerable function \cite{Ferrie2008}. Malware like \cite{Brulez} exploited these vulnerabilities would cause the "bluescreen of death. \textit{SoftICE} is no longer supported and the \textit{dll} file that caused the vulnerability in \textit{OllyDBG} is now fixed. But the idea still remains, if a vulnerability within a specific debugger is discovered, exploiting it would be a potent anti-debugging technique. Table 1 summarizes the results of our study. 

\paragraph{\textbf{Fileless malware}}\label{fileless malware}
Fileless malware, non-malware and occasionally called Advanced Volatile Attack (AVT) attacks, are the latest trend in the evolution of malware. \cite{patten2017evolution}. In contrast to all prior existence of malware, fileless malware requires no file to operate and they purely reside in memory and take advantage of existing system tools e.g. \textit{powershell} \cite{mansfield2017fileless}. 

The purpose of such attacks is to make the forensics much harder. To analyze a malware is to analyze its executable; and in fileless malware, there is no executable to begin with. In some cases such as SamSam \cite{Malwarebytes2018} the only way to just retrieve a sample for analysis would be to catch the attack taking place live. These attacks inherently are not easy to conduct, but, with the help of exploit-kits are more readily available. A 2018 report by McAfee shows a 432\% increase of fileless malware in 2017 \cite{Bassett2018} and projected to constitute 35\% of attacks in 2018 \cite{LarryPonemon2018}.

With the aid of debuggers, the analyst can overcome many hurdles and limitations of static analysis in the manual dynamic analysis. However, new trends and real world scenarios in which the vendors face thousands of new malware samples daily demands a more agile approach which is beyond the capabilities of manual dynamic analysis. In the next section, we discuss the emergence of automated dynamic analysis approach and Sandboxes as a response to these challenges and elaborate malware’s tactics to thwart them. Table 1 summarizes our survey of manual dynamic evasion techniques. 
\begin{table}
  \caption{Classification and comparison of malware anti-debugging techniques}
  \includegraphics[scale=0.75]{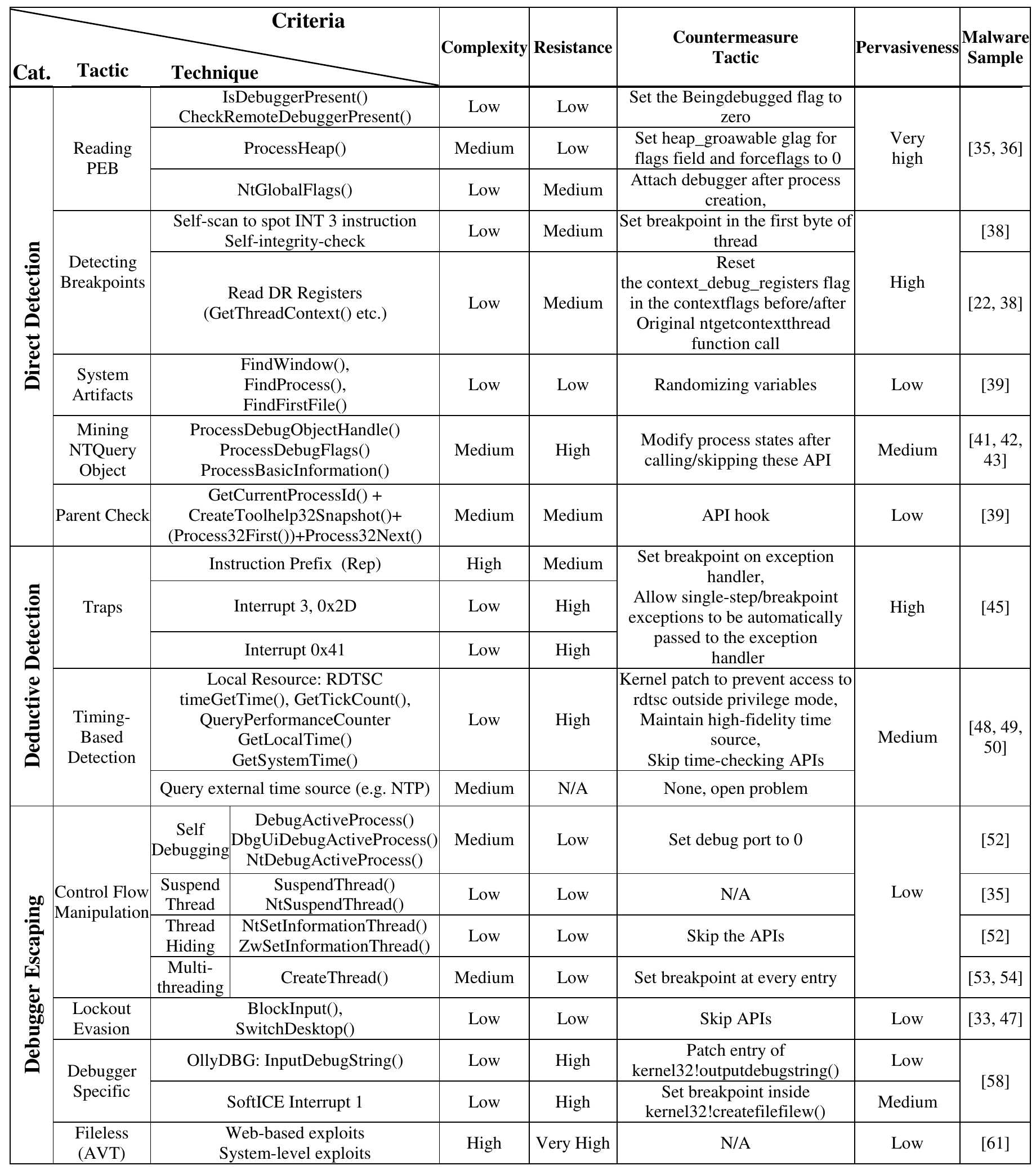}
\end{table}

\subsection{A Discussion}
In table 1 we summarize our survey on malware manual dynamic analysis evasion. In this table, we compare the techniques based on three criteria i.e. complexity of implementation, resistance against manual dynamic analysis, and pervasiveness. In addition, in the example column we provide malware samples that incorporate the corresponding techniques. 

Manual dynamic analysis evasion, or anti-debugging is as old as the existence of software. What we have observed in the last decade, attest that incorporation of anti-debugging is prevalent and has been steadily increasing. Authors in \cite{chen2016advanced} demonstrate that on average, more than 70 of malware samples utilize anti-debugging techniques. The majority of these techniques are the ones we have covered in this section. 
In addition, we have noticed that besides the very pervasive techniques such as \textit{IsDebuggerPresent}, malware are also employing more timing-based techniques.

Another emerging trend in malware industry, is fileless malware. Such malware as in the case of Rozena \cite{Go2018} leverages exploits to execute malicious commands right from the memory. Given the difficulties of acquiring samples in post-infection phase and the fact that they reside merely in memory, perhaps we can conclude that fileless or AVT attack is the most effective manual dynamic analysis evasion tactic.

\section{Automated Dynamic Analysis Evasion}
Although effective, the manual dynamic analysis suffers a critical limitation, that is, time. A 2018 statistics provided by \textit{McAfee} reports on receiving more than 600K new samples each day \cite{Labs2018}. Analyzing this massive number of malware samples calls for a far more agile approach. This demand led to a new paradigm of analysis which we referred to as automated dynamic analysis. Sandbox, is the representative technology for this paradigm. In this section, we will have a brief introduction to sandboxes and further propose a classification and comparison of malware evasion tactics.

\subsection{An overview of malware sandboxes}
The concept behind a malware automated dynamic analysis system is to capture the suspicious program in a controlled and contained testing environment called sandbox, where its behavior in runtime can be closely studied and analyzed. Initially, sandboxes were employed as a part of the manual malware analysis. But today, they are playing their roles as the core of automated detection process \cite{Kruegel2014}. Sandboxes are built in different ways. To better grasp the evasion tactics, and depict how they stand against different sandbox technologies, first, we must have a sense of how they are made.

\subsubsection{Virtualization-based sandboxes}
A virtual machine (VM) according to Goldberg \cite{Goldberg1974}, is “an efficient, isolated duplicate of the real machine.” The hypervisor or virtual machine manager (\textit{VMM}) is in charge of managing and mediating programs’ access requests to the underlying hardware. In other words, every virtual machine atop the \textit{VMM}, in order to access the hardware, must first pass through the hypervisor. There are a couple of ways to implement sandboxes based on virtualization. One way is to weave the analysis tools directly into the hypervisor as in the case of ether \cite{Dinaburga}. The other approach would be to embed the analysis tools (e.g. installing hooks) within the virtual machine that runs the malware sample. Instances of this design are: \textit{Norman} sandbox \cite{norman}, \textit{CWsandbox} \cite{Willems2007}. Both of the methods inherently leak subtle cues which a malware could pick on to detect the presence of a sandbox. In the latter case, for instance, the \textit{VMM} has to provide the required information to the analysis VM which means whenever a sensitive system call is being made by the malware, the \textit{VMM} has to pass the control to the analysis tools inside the VM. Challenges of performing these procedures without leakage are profound which we will elaborate accordingly.  

\subsubsection{Emulation-based sandboxes}
An emulator is a software that simulates a functionality or a piece of hardware \cite{Egele2012}. An emulation-based sandbox can be achieved through different designs. One would be to simulate the necessary OS functions and APIs. Another approach is the simulation of CPU and memory and is the case for many anti-virus products \cite{Egele2012}. Simulation of I/O in addition to memory and CPU is what in literature is referred to as the full system simulation. The eminent features of emulation-based sandboxes are the great flexibility and detailed visibility of malware inner workings (introspection) they offer. Especially with the full system emulation, the behavior of the program under inspection (PUI) could be studied with minute details.

\subsubsection{Bare-metal sandboxes}
Recent evolved and perplexing evasion tactics, employed by sophisticated malware demands a new paradigm of analysis. The emerging idea is to execute the malware in several different analysis environments simultaneously with the assumption that any deviation in behavior is a potential indication of malicious intents \cite{kirat2014barecloud}. The feasibility of this idea requires a reference system in which the malware is analyzed without utilization of any detectable component and the ideal choice would be a bare-metal environment equal to a real production system in terms of transparency. There have been several products in this vein e.g. \textit{Barebox}, bare cloud etc. \cite{kirat2014barecloud,Willems, kirat2011barebox}. 
Along with the merits that each design offers, subtle flaws or specific working principles that malware exploit to forge their evasion tactic.

\subsection{Proposed Classification of Automated Dynamic Analysis (sandbox) Evasion tactics}
 If the malware achieves one specific goal, it can triumphantly evade the sandbox. This goal is to behave nicely or refrain from executing its malicious payload so long as it resides within the sandbox. This strategy capitalizes on two facts. The first is that due to a massive number of malware samples and limitation of resources, sandboxes allot a specific limited time to the analysis of a sample. The second stems from an inherent limitation of dynamic analysis. In dynamic analysis since the “runtime behavior” and “execution” is being inspected, only the execution path is visible to the inspector (sandbox). Thus, if a malware portrays no malevolent behavior while under examination, the sandbox flags it as benign.
 
 We propose to classify automated dynamic analysis evasion tactics under two categories i.e. Detection-Dependent evasion and Detection-Independent evasion. We will elaborate on these tactics along with several techniques under each tactic. In addition, we briefly hint at the way sandboxes try to defeat these tactics. To provide a more coherent overview of these evasion techniques, we employ three criteria as follows which would serve as the basis for our comparison:

\begin{itemize}
    \item \textbf{Complexity:} The difficulty of implementing an evasion tactic. Our touchstone for evaluating complexity are  lines of code (\textit{LOC}), and more importantly the challenges which differ per tactic. We will discuss these challenges in the elaboration on the tactics
    \item \textbf{Pervasiveness:} This criterion captures the relative prevalence of a particular tactic. Based on our observations or security reports \cite{Labs2018,McAfee2017}, and security reports, we try to provide a relative prevalence for each tactic. 
    \item \textbf{Efficacy-Level:} Each tactic have an efficacy level. This demonstrates against which type of sandbox that particular tactic is effective. 
\end{itemize}

\subsubsection{Detection-Dependent Evasion}

\paragraph{Fingerprinting}
Fingerprinting is a tactic pursued by malware to detect the presence of sandboxes by looking for environmental artifacts or signs that could reveal the indications of a virtual/emulated machine. These signs can rangefrom device drivers, overt files on disk and registry keys to discrepancies emulated/virtualized processors. Indications of \textit{VM} or an emulation are scattered in different levels \cite{Raffetseder,Lau2010,Holz,Kapravelosa}. It is noteworthy to mention that initially, many sandboxes such as Norman \cite{norman} were developed upon \textit{VMs}. Thus, in literature, you may still observe the terms sandbox and \textit{VMs} being used interchangeably to imply a contained analysis environment. The very same fact is also the reason for the advent of techniques referred to as anti-VM, suggesting that detection of a virtual machine would potentially mean an analysis environment. 
Different studies have been conducted to uncover the levels at which sandboxes leave their marks \cite{Chen,Yokoyama}. These levels are:

\begin{itemize}
    \item \textbf{Hardware:} Devices and drivers are overt artifacts that malware might look for to identify its environment. In the case of devices, \textit{VMs} often emulate devices that can be readily detected as in the case of Reptile malware \cite{reptile2012}. This ranges from obvious footprints such as the \textit{VMWare} Ethernet device with its identifiable manufacturer prefix, to more subtle marks. Moreover, specific drivers are employed by VMs to interact properly with the host OS. These drivers are other indications of analysis environment for malware. For instance in the path: \verb|C:\Windows\System32\Drivers| exist such signs that could expose \textit{VMWare}, \textit{VirtualBox} etc. (e.g. \textit{Vmmouse.sys}, \textit{vm3dgl.dll}, \textit{VMToolsHook.dll} etc.) \cite{Assor2016}.
    \item \textbf{Execution Environment:} A malware inside sandbox experiences subtle differences in the environment within which they are executed. Kernel space memory values, for instance, are different between a sandbox and native system that can be detected by malware as in the cases of \textit{Agobot} and \textit{Storm} Trojans \cite{Sophos2015,Symantec2007}. Artifacts of this level manifest themselves either in memory or execution. As in the case of memory artifacts, for instance, to allow inspection and control between host and guest OS, \textit{VMWare} creates a channel between them (“\textit{ComCHannel}”). Virtual PC hooks work in the same way \cite{Kruegel}.
    \item \textbf{Application:} When an analysis application is used in an environment, its presence is usually disclosed due to the artifacts of installation and execution which can be picked up by inquisitive malware such as \textit{Rbot} \cite{Microsoft2017} or \textit{Phatbot} \cite{Microsoft2006}. Even if not executed, evidence of the analysis tools might be residing on the disc, registry keys etc. which could be readily found by malware especially if they contain well-known file name or locations e.g. \verb|SYSTEM\CurrentControlSet\Control\VirtualDeviceDrivers|. In a similar instance, if the names are not altered or the corresponding processes concealed, they can be enumerated by malware with little trouble. For instance, VMtools.exe, \textit{Vmwareuser.exe} or \textit{vboxservice.exe} if queried by malware, are vivid indicators of a virtual machine and analysis environment.
    \item \textbf{Behavior:} Perhaps, the most troublesome of all, for anti-malware vendors to conceal, are the leakages caused by imperfect virtualization/emulation, or characteristics that are innate to such environments. Of particular interest are discrepancies of behavior between an emulated CPU and a physical one \cite{Raffetseder}. Performance of an application running under emulation is inferior to a real system. This performance penalty stems from translations or interceptions that has to be carried out in an emulation/virtualization. Such discrepancies can be disclosed through a diverse set of timing attacks \cite{Franklin2008,Pek,Garfinkel}.  Computing absolute performance for fingerprinting is an arduous task due to the diversity of hardware configuration. 
    
    An alternative for malware is to calculate the relative performance \cite{Raffetseder}. Pursuing this method, malware compares performance ratio of two or more operations on the same system. If the measurement varies significantly among production and emulation systems, there’s a high chance of sandbox presence. Another interesting way of exploiting the limitations of emulation is observing and comparing the effects of caching on emulated and real environments. In this technique, a function is executed a number of times. As expected, the first run must to be the slowest. The same test is performed again, this time, in absence of caching. Timing analysis depicts the effectiveness of the employed caching. Simulation of processor cache is a complex task and emulators may not support it, the result of which is exposure of emulation in the aforementioned test \cite{Raffetseder}. Another noteworthy set of techniques are the red pill tests \cite{Paleari}. An inevitable behavioral artifact sandboxes share, are the imperfect simulation of CPU and residing instruction bugs. If a malware specifically finds such bugs and if during the execution, it notices a mishandled instruction it will suspect the presence of the sandbox.
    \item \textbf{Network} In addition to the previous levels that were suggested by Chen et al. \cite{Chen}, malware can also probe the network in pursuit of sandbox's  marks. These marks are manifested in many forms such as known fixed IP addresses, \cite{yoshioka2011your}, limitations pertaining to the sandboxes that prevent or emulate the internet access \cite{Blackthorne,spensky2016phi} or extremely fast internet connection \cite{Dolan-Gavitt2010}. One technique for instance, proposed in \cite{yoshioka2011your} is detecting sandboxes based their known IP addresses. which is acquired in an earlier attack through a decoy malware. 
\end{itemize}
Fingerprinting tactic was the initial endeavor of malware authors to detect and evade sandboxes and as we observed the techniques have evolved significantly. In countering fingerprinting, most solutions are reactive, namely, the fingerprinting technique must first be disclosed, then the corresponding counter evasion will be provisioned. Collectively, fingerprinting tactic is still the dominant approach to evade sandboxes. \cite{McAfee2017}.

\paragraph{Reverse Turing Test}
 The second tactic that aims at detection, is checking for human interaction with the system. This tactic capitalizes on the fact that sandboxes are automated machines with no human or operator directly interacting with them. Thus, if a malware does not observe any human interaction, it presumes to be in a sandbox. Such tactic is referred to as Reverse Turing Test since a machine is trying to distinguish between human or AI. This tactic can be carried out through various techniques \cite{Kapravelosa,Singh2017,Falliere2011,Dolan-Gavitt2010,AbhishekSingh2012}. For instance, the \textit{UpClicker} \cite{AbhishekSingh2012} or a more advanced one, \textit{BaneChant} \cite{Hwa2013} await the mouse left-click to detect human interaction \cite{SaiOmkarVashisht2014}. 
 In a large portion of reverse Turing test techniques, malware looks for last user’s inputs. To do so, malware often leverages a combination of \textit{GetTickCount()} and \textit{GetLastInputInfo()} functions to compute the idle time of the user. To test whether it’s running on a real system, the malware waits indefinitely for any form of user input. On a real system eventually, a key would be pressed or mouse would be moved by the user. If that occurs for a specific number of times, malware executes its malicious payload. To counter this tactic of evasion, simulating human behavior might seem intuitive but it might be counterproductive. One reason is that digitally-generated human behavior can also be detected \cite{ArunpreetSingh2014}. The second reason is that limitation of designing such reversal Turing test seems to be merely a function of imagination. To demonstrate our point, consider another technique in which the malware waits for the user to scroll to the second page of a Rich Text Format (RTF) before it executes the malicious payload. Or In another approach, using Windows function \textit{GetCursorPos()}, which holds the position of system’s cursor, the malware checks the cursor movement speed between instructions; if it exceeds a specific threshold, it would imply that the movement is too fast to be human-generated and malware ceases to operate \cite{SaiOmkarVashisht2014}; or another more recent technique which relies on seeking wear and tear signs of a production system \cite{Miramirkhani}. The seemingly endless possibilities for designing reversal Turing tests makes it a puzzling challenge for sandboxes to counter. This tactic is becoming more prevalent, but still not as pervasive as fingerprinting \cite{McAfee2017}.

\paragraph{Targeted}
The third tactic of the detection-dependent category is targeted detection. This tactic is slightly different from the previous ones in that instead of striving for detecting or evading a sandbox directly, the malware fingerprints the environment to verify if the host is precisely the intended (targeted) machine. In other words, malware looks for its target, not sandbox. This tactic can be employed following different routes: 
\begin{itemize}
    \item \textbf{Environmentally-targeted:} \textit{Stuxnet} is known to be the first Cyber weapon that incorporated targeting tactic as a portion of its evasion tactics \cite{Falliere2011}. \textit{Stuxnet} explicitly looked for the presence of a specific industrial control system and would remain dormant otherwise. Depending on the infection approach, APT attacks follow different routes. 
    \item \textbf{Individually Targeted:} In the case of \textit{Stuxnet}, the strategy was to keep probing victims (a wormy behavior) until the target was reached. Other classes of APT include the attacks such as \textit{darkhotel} \cite{GReAT2014}, in which the infection is conducted through spear phishing (i.e. directly aimed at the target).In other words, the attackers make sure that their malicious code is directly delivered to their target.
    \item \textbf{Environment-dependent Encryption:} Such targeted malware, have an encrypted payload the decryption of which, is subjected to a key that is derived from its victim's environment. The key might be a hardware serial number, specific environmental settings, etc. You can refer to \cite{Morrow2016} for a thorough discussion of the topic.
\end{itemize}

\subsubsection{Detection-Independent Evasion}
The major difference of this category of tactics is that they do not rely on detecting the target environment, and their evasion tactic is independent of target system which relieves them from having to employ sophisticated detection techniques. Consequently, efforts directed at achieving more transparency has no effects on these tactics. In the following we elaborate on the tactics of this category and several techniques to deploy them. 

\paragraph{Stalling}
This tactic of analysis evasion capitalizes on the fact that sandboxes allot a limited amount of time to the analysis of each sample. A malware has to simply postpone its malicious activity to the post-analysis stage. To this end, malware authors have come up with the idea of stalling \cite{Kruegela} which can be achieved through a diverse set of techniques that range from a simple call to sleep() function to more sophisticated ones. Here we examine some of the known major stalling techniques.
\begin{itemize}
\item \textbf{Simple Sleep:} Sleeping is the simplest form of stalling and just as simple to defeat. The idea was to remain inactive for n minutes in order for the sandbox inspection to timeout before observing any malicious activity. After being released to the network, the malware would execute the malicious payload. The infamous \textit{DUQU} \cite{Bencsath2012} exhibits such technique as one of its prerequisites for ignition. It requires that the system remain idle for at least 10 minutes \cite{Schiffman2010}.  Another example would be the \textit{Khelios} botnet \cite{MalwareTech2015}. A new variant of \textit{Khelios} sample found in 2013 (Called Nap) calls the \textit{SleepEx()} API with a timeout of 10 minutes. This delay in execution outruns the sandbox analysis timeout which is followed by achieving the harmless flag. To counter such techniques, sandboxes came up with the simple idea of accelerating the time (called sleep patching). Although seemingly logical, sleep patching has unpredicted effects. An interesting side effect of sleep patching was observed in specific malware samples where the acceleration actually lead to inactivity of malware instances that wait for human interaction within a predefined period of time \cite{ArunpreetSingh2014}. Despite its side effects, sleep patching turned out to be effective in some cases and malware authors came up with more advanced techniques as a response.
\item \textbf{Advanced Sleep:} New malware instances such as \textit{Pafish} \cite{pafish2014a} were found to opt for more advanced sleeping tricks. They detect sleep-patching using the \textit{rdtsc}  instruction in combination with Sleep() to check the acceleration of execution.
\item \textbf{Code stalling}: While "sleeps", delay the execution, in code stalling the malware, opts for executing irrelevant and time-consuming benign instruction to avoid raising any suspicions from the sandbox \cite{ArunpreetSingh2014}. \textit{Rombertik} \cite{BenBaker2015} is a spyware aimed at stealing confidential data. To confuse sandboxes, it writes approximately 960 million bytes of random data to memory \cite{Lindorfer}. The hurdles this technique impose on sandboxes are two-fold.  The first one is the inability of the sandbox to suspect stalling as the sample is running actively. The second one, is that this excessive writing would overwhelm the tracking tools \cite{BenBaker2015}. In another striking code stalling technique, the malware encrypts the payload using a weak encryption key and brute-forces it during the execution \cite{kirat2014barecloud}.
\end{itemize}

\paragraph{Trigger-based}
The second tactic that does not rely on detection is a trigger-base tactic or more traditionally logic bombs. As we have already noted, the basis of evading sandboxes is to simply refrain from exhibiting the malicious behavior so long as they are in the sandbox. Another tactic for remaining dormant would be to wait for a trigger. There are many environmental variables that can serve as malware triggers ranging from system date to a specially crafted network instruction. For instance, \textit{MyDoom} \cite{McAfee2007} is triggered on specific dates and performs \textit{DDoS} attacks, or some key-loggers only log keystrokes for particular websites; and finally, \textit{DDoS} zombies which are only activated when given the proper command \cite{Barlow2000}. In the literature, this behavior is referred to as trigger-based behavior \cite{Brumley2008}. In the following there are a number of triggers embedded within malware as a protection against sandboxes.

\begin{itemize}
\item \textbf{Keystroke-based:} The malware  gets triggered if it notices a specific keyword, for instance, the name of an app or the title of a window,\cite{Roberts2004}. What occurs when the trigger happens, depends on the purpose and context. For example, the malware might initiate logging keystrokes. 
\item \textbf{System time:} In this case, the system date or time would serve as the trigger \cite{Roberts2004,Symantec2000,McAfee2003}
\item \textbf{Network Inputs:} A portion of malware are triggered when they receive certain inputs from the network as in the case of Tribal flood network \cite{Branco2012}.
\item \textbf{Covert trigger-based:} Previous techniques often presuppose lack of code inspection to remain undetected (as it is often the case for compiled malware). However, there have been advances regarding automatic detection of such program routes e.g. State-of-the-art covert trigger-based techniques are being devised as a response to automated approaches that aim at detecting such triggers instruction. These techniques utilize instruction-level stenography to hide the malicious code from the disassemblers. In addition, they implement trigger-based bugs to provision stealthy control transfer which makes it difficult for dynamic analysis to discover proper triggers \cite{Falliere2011} return values from system calls are other instances of malware triggers.     
\end{itemize}

Trigger-based tactic initially was not utilized to evade sandboxes and are still relatively seen in the wild. Finding these triggers is often pursued through path exploration approaches e.g. symbolic execution with the goal of finding and traversing all the conditional branches in an automated manner. These approaches require significant levels of resources and are still below a fair level of efficiency. 
 
\paragraph{Fileless Malware} In \ref{fileless malware} we discussed the fileless malware. In addition to the rigorous resistance against manual analysis, fileless malware is profoundly adept at evading security mechanisms. A major difference of this tactic is the most often the malware is not subjected to the analysis environment in the first place. This is an inherent outcome of this tactic. The techniques that fileless malware utilizes are similar to drive-by attacks \cite{le2013anatomy}. Generally, fileless malware exploits a vulnerability of target system (OS, Browser, Browser plugins, etc.) and injects its malicious code directly into memory. The new trend of fileless malware uses windows \textit{PowerShell} to carryout its task. As mentioned earlier, this attack is on the rise, and defending against it is of great complexity.  

\begin{table}
  \caption{Classification and comparison of malware sandbox evasion techniques}
  \includegraphics[scale=0.65]{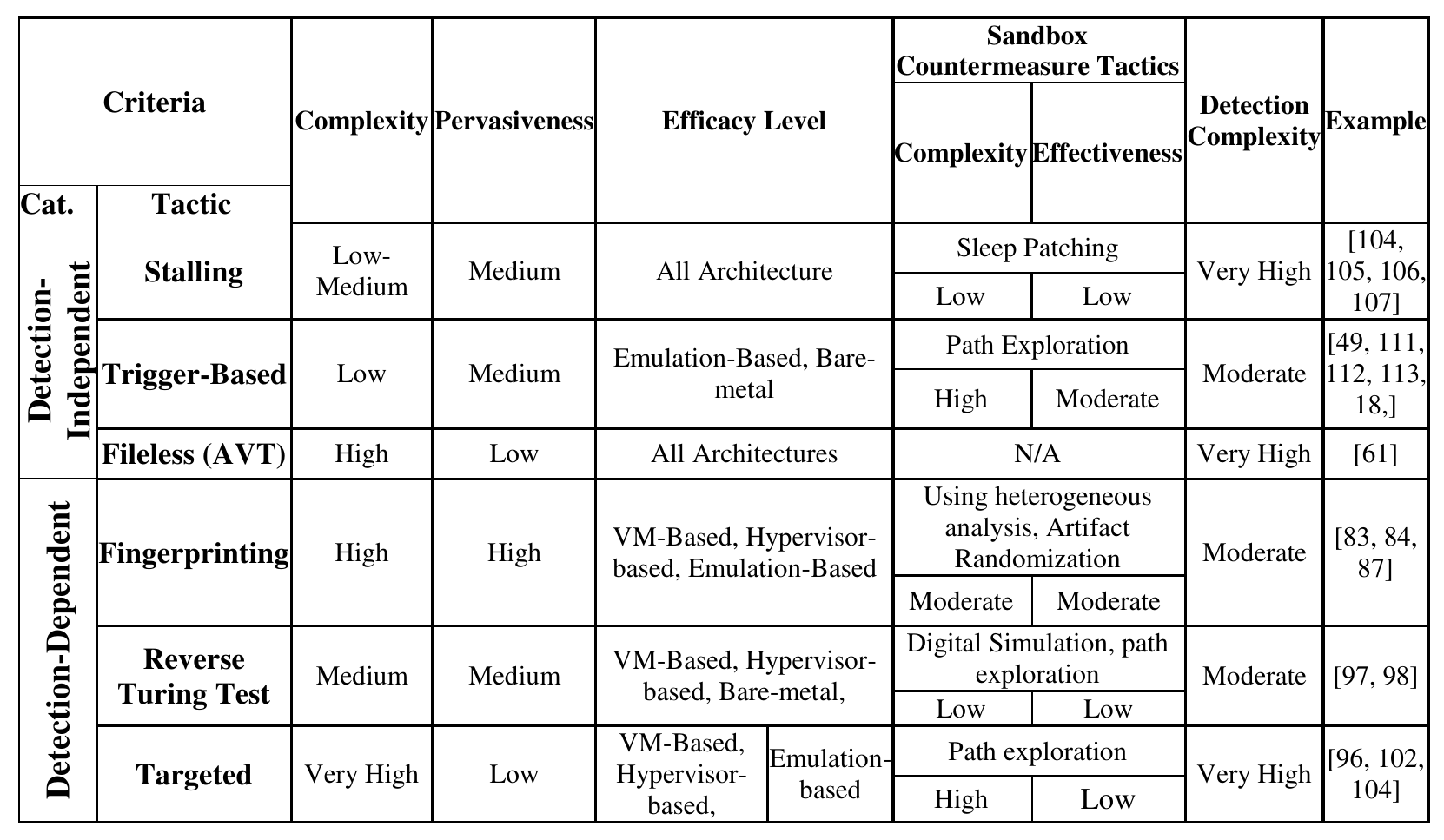}
\end{table}

\subsection{Discussion}
In table 2 we summarize our survey on malware automated dynamic analysis evasion. In this table, we compare the techniques based on four criteria i.e. complexity of implementation, pervasiveness, efficacy level and detection complexity. In addition, in the example column we provide malware samples that incorporate the corresponding techniques. Moreover, under Sandbox countermeasure tactics, we hint on how the defensive side strives to counter the evasion tactics and how complex/effective these countermeasures are.

In this section we provided our survey and classification on malware automated dynamic analysis evasion. Our observations, in line with studies such as \cite{McAfee2017, chen2016advanced, oyama2018trends} notes several trends.

Fingerprinting by far, is the most pervasive tactic and has proven to be an effective technique if executed with precision. Aside from other researches and our observations, another proof regarding the dominance of fingerprinting tactic is the trend of defensive research in the last decade and how they are evolving \cite{vasudevan2006cobra, moser2007exploring, Brumley2008, Dinaburga, nguyen2009mavmm, kirat2011barebox, yan2012v2e, royal2012entrapment, zhang2013spectre, kirat2014barecloud, zhang2015using, leach2016towards, spensky2016phi, zhang2018towards}. Examining these researches, we notice that the great majority of them are pursuing more transparent system the reason of which is to be more resilient against fingerprinting and more broadly, detection-dependent tactics.

We see two reasons behind the pervasiveness of fingerprinting tactic. The first reason is the immensity of possibilities for fingerprinting scenarios. In each level of the system (application, network, behavior, etc.) there are always new hints of identity that a malware can look for. 

The second reason, is the countermeasure strategies against fingerprinting tactic. There are two major strategies in this regard. The first one, is the reactive approach. Namely, for the sandbox to neutralize malware attempts for fingerprinting, it has to know what constitutes fingerprinting in the first place. The second strategy is to aim for more transparent systems either following the approach of single-stand analysis systems e.g. \cite{Dinaburga} or heterogeneous analysis i.e. bare-metal analysis \cite{kirat2011barebox, kirat2014barecloud}. In retrospect to the cases we cited in this section, we know that both strategies are still vulnerable to zero-day fingerprinting techniques. 

Although attempts for achieving more transparency has raised the bar for fingerprinting, new reversal Turing test tactic employed by malware virtually foils most of the efforts. Reverse Turing test tactics again, pushes the defenders to defend with the reactive approach. 

On the other hand we are noticing new trends of malware attack employing detection-independent tactics. Most notably stalling \cite{oyama2018trends} and fileless tactics \cite{Bassett2018}. The majority of the efforts that has been put into developing more transparent systems, has raised the bar for detection-dependent tactics. That is one reason that has motivated malware authors to adopt more detection-independent tactics.

Unbound fingerprinting possibilities, reactive approaches, vulnerability of transparent systems to zero-day detection-dependent tactics (fingerprinting, targeted, reverse Turing), vulnerability of transparency-seeking strategies to detection-independent tactics, all seem to call for more generic and effective defensive strategies.

A promising approach that can potentially counter most of the evasion tactics are path exploration techniques. Every malware will expose its malicious payload if its specific conditions are satisfied. Path exploration techniques potentially can  trigger those conditions and expose the malicious behavior.

\section{Conclusion and Future Directions}
In this paper, we conducted a surveyed on  the subject of malware analysis evasion techniques, and proposed classifications for both modes of manual and automated analysis. 

For the automated dynamic analysis, we defined two major categories of evasion. Detection-dependent and detection-independent. On the defensive side, defenders are pursuing 4 major strategies. Reactive approaches, more transparent analysis systems, bare-metal analysis, and several endeavors towards more generic approaches, namely, path exploration. The first three defensive strategies can hope for being effective merely against detection-dependent evasion tactics and loose their effectiveness against detection-independent tactics. As for the detection-independent tactic, we need more generic approaches such as the path exploration methods. 

Parallel to endeavors towards path exploration methods, we advocate automated fingerprinting generations such as MORPHEUS \cite{Jing}, and Red Pills \cite{Paleari}. These solutions would not be a generic response, but they would significantly raise the bar for malware pursuing fingerprinting tactic.

Another line of inquiry that we suggest is analyzing the effectiveness of using malware evasion tactics against itself; for the case of detection-dependant tactics, a malware strives relentlessly to detect a Sandbox and refrain from execution. What would be the result of a program, installed on a production system that would impersonate a Sandbox. How efficacious it would be in deterring malware and if executed adroitly, how significantly such approach  would raise the bar for devising future evasion attempts.

\section{Acknowledgments}This work is supported by APA research center (http://apa.aut.ac.ir)
at Amirkabir University of Technology, Tehran, Iran. In addition, would like to expand our gratitude to Prof. Eric Filiol of ESIEA (C + V)O lab for his support and insights on this paper.

\bibliographystyle{unsrt}
\bibliography{09_bibliography}

\end{document}